\documentclass[12pt,draftclsnofoot, perreview, onecolumn]{IEEEtran}
\usepackage{balance}
\usepackage{color}
\usepackage{url}
\usepackage{threeparttable}
\usepackage{amsfonts}
\usepackage{amsmath}
\usepackage{mathrsfs}
\usepackage{multirow}
\usepackage{amssymb}
\usepackage{mdwmath}
\usepackage{bm}
\usepackage{ifpdf}
\usepackage{setspace}
\usepackage{booktabs}
\usepackage{cite}

\ifCLASSINFOpdf
  \usepackage[pdftex]{graphicx}
  \graphicspath{{../pdf/}{../jpeg/}}
  \DeclareGraphicsExtensions{.pdf,.jpeg,.png}
\else
  \usepackage[dvips]{graphicx}
  \graphicspath{{../eps/}}
  \DeclareGraphicsExtensions{.eps}
\fi

\ifCLASSOPTIONcompsoc
  \usepackage[tight,normalsize,sf,SF]{subfigure}
\else
  \usepackage[tight,footnotesize]{subfigure}
\fi

\usepackage{pict2e}
\makeatletter
\newcommand*{\bigboxplus}{\DOTSB\mathop{\mathpalette\big@boxplus\relax}\slimits@}

\newcommand{\big@boxplus}[2]{%
  \vcenter{%
    \m@th\bigbox@thickness{#1}%
    \sbox\z@{$#1\bigoplus$}%
    \dimen@=\ht\z@ \advance\dimen@\dp\z@
    \hbox{%
      \setlength{\unitlength}{\dimen@}%
      \begin{picture}(1,1)
      \polyline(0.1,0.1)(0.9,0.1)(0.9,0.9)(0.1,0.9)(0.1,0.1)(0.5,0.1)
      \polyline(0.5,0.1)(0.5,0.9)
      \polyline(0.1,0.5)(0.9,0.5)
      \end{picture}%
    }%
  }%
}

\newcommand{\bigbox@thickness}[1]{%
  \ifx#1\displaystyle
    \linethickness{0.2ex}%
  \else
    \ifx#1\textstyle
      \linethickness{0.16ex}%
    \else
      \ifx#1\scriptstyle
        \linethickness{0.12ex}%
      \else
        \linethickness{0.1ex}%
      \fi
    \fi
  \fi
}
\makeatother

\newtheorem{algorithm}{\textbf{Algorithm}}
\newtheorem{example}{\textbf{Example}}

\ifCLASSOPTIONonecolumn
\newcommand{\figwidth}{0.65\textwidth}

\fi

\ifCLASSOPTIONtwocolumn
\newcommand{\figwidth}{0.48\textwidth}

\fi

\begin{document}
\title{Twisted-Pair Superposition Transmission}

\author{Suihua Cai and Xiao Ma,~\IEEEmembership{Member,~IEEE}
\thanks{This work was supported by the Science and Technology Planning Project of Guangdong Province (2018B010114001), the National Key Research and Development Program of China~(No.~2017YFE0112600), the NSF of China~(No.~61771499 and No.~61971454), and  China Postdoctoral Science Foundation~(2020TQ0371). This work was presented in part at 2020 IEEE International Symposium on Information Theory. \emph{(Corresponding author: Xiao Ma.)}}
\thanks{The authors are with the School of Data and Computer Science and Guangdong Key Laboratory of Information Security Technology, Sun Yat-sen University, Guangzhou 510006, China (e-mail: caish23@mail.sysu.edu.cn, maxiao@mail.sysu.edu.cn).}
}

\maketitle

\begin{abstract}
We propose in this paper a new coding scheme called twisted-pair superposition transmission~(TPST).
The encoding is to ``mix together'' a pair of basic codes by superposition, while the decoding can be implemented as a successive cancellation list decoding algorithm.
The most significant features of the TPST code are its predictable performance that can be estimated numerically from the basic codes and its flexible construction in the sense that it can be easily adapted to different coding rates.
To construct good TPST codes in the finite length regime, we propose two design approaches -- rate allocation and partial superposition.
By taking tail-biting convolutional codes~(TBCC) as basic codes, we show by numerical results that the TPST codes can have near-capacity performance in the short length regime.

\end{abstract}


\section{Introduction}

With the development of modern communications, ultra-reliable and low latency communication~(URLLC), one of the main scenarios defined in the 5G network, has attracted more and more attention.
To meet the stringent requirements of low latency, the channel codes for URLLC should be designed for short block length $n$, say $n<1000$~\cite{shirvanimoghaddam2019short}.

Theoretically, several bounds on the channel capacity in the finite length regime have been presented in~\cite{polyanskiy2010channel}, where the decoding performance in terms of frame error rate~(FER) for the maximum likelihood~(ML) decoding is upper bounded by the random coding union~(RCU) bound and lower bounded by the meta-converse~(MC) bound.
To approach the performance bounds, coding schemes tailored to the transmission of short blocks were surveyed in~\cite{coskun2019efficient}, where BCH codes, tail-biting convolutional codes~(TBCCs), polar codes, and low-density parity-check~(LDPC) codes are compared and shown to be potentially applicable for low latency communications.
It is believed that list decoding algorithms such as order statistic decoding~(OSD)~\cite{Fossorier1995soft} can be used to attain a near-ML performance for codes such as BCH codes and Reed-Muller codes with length $n\leqslant 128$.

Another efficient list decoding algorithm is the list Viterbi algorithm~\cite{seshadri1994list} for trellis codes such as TBCCs, which generates a list of candidate codewords associated with the highest likelihoods.
It is shown in~\cite{stahl1999optimal} that, TBCCs with a large encoding memory can attain a near-capacity performance with short and medium length.
However, it suffers from a high decoding complexity.
To reduce the decoding complexity, TBCCs are designed by concatenating with cyclic redundancy checks~(CRCs)~\cite{Liang2019list}.
The TBCC with CRC has been adopted in the Long-Term Evolution (LTE) standard, where the outer concatenated CRC can be viewed as an error detector for identifying the correct candidate codeword in the decoding list~\cite{kim2018anew}.

Similarly, polar codes decoded by the successive cancelation list~(SCL) decoding can be significantly improved by concatenating with CRCs~\cite{Niu2012CRC,Tal2015list}.
In his Shannon Lecture, Ar{\i}kan presented the  polarization-adjusted convolutional~(PAC) code~\cite{Arikan2019sequential}, a new polar code construction whose CRC precoding is replaced by rate-one convolutional precoding.
The performance is further improved to approach the RCU bound by sequential decoding of the PAC codes.
From the perspective of polar coding with dynamically frozen bits, the list decoding algorithm of PAC codes was investigated in~\cite{Yao2020list}, which is shown to closely match the performance of the sequential decoder.

In this paper, we propose a new coding scheme called twisted-pair superposition transmission~(TPST), which is constructed by superimposing together a pair of basic codes in a twisted manner~\cite{Cai2020twisted}.
We propose a successive cancellation list decoding algorithm for the TPST codes, which can be early terminated by a preset threshold on the empirical divergence functions~(EDF) to trade off performance with decoding complexity.
The SCL decoding of TPST is based on the efficient list decoding of the basic codes, where the correct candidate codeword in the decoding list is distinguished by employing a typicality-based statistical learning aided decoding algorithm~\cite{ma2019statistical}.
We derive lower bounds for the two layers of TPST, which can be used to predict the decoding performance and  to show the near-ML performance of the proposed SCL decoding algorithm.
To optimize the performance of the TPST codes, we present two design approaches -- rate allocation and partial superposition.
Numerical simulation results show that, the constructed TPST codes have near-capacity performance in the short length regime, suggesting that the TPST codes may find applications in low latency communications as an attractive candidate.

The rest of this paper is organized as follows. The encoding and decoding algorithm of the TPST code are introduced in Section~\ref{sec:TPST}. The performance of the TPST code is analyzed by deriving bounds for FER in Section~\ref{sec:performance}, where also presented are the complexity analysis and an early termination strategy.
Design approaches for the construction of good TPST codes with short length are presented in Section~\ref{sec:design}. Numerical results are presented in Section~\ref{sec:simulation}. Finally, Section~\ref{sec:conclusion} summarizes this paper.

\section{Twisted-pair Superposition Transmission}\label{sec:TPST}
\subsection{Encoding}
\begin{figure}
  \centering
  \includegraphics[width=\figwidth]{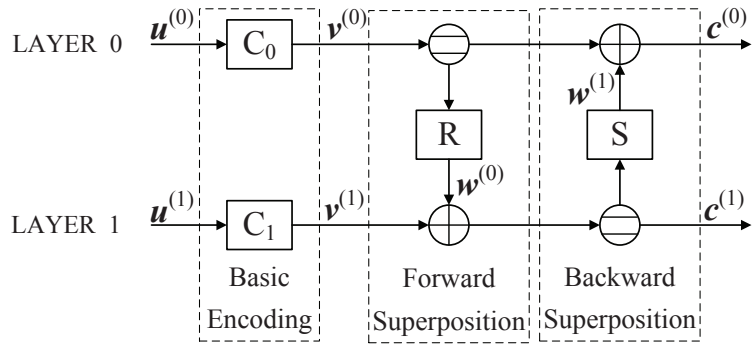}
  \caption{Encoding structure of a TPST code. $C_0$ and $C_1$ are the basic codes with length $n$ of the two layers, $R$ is an $n\times n$ binary random matrix, and $S$ is the $n\times n$ binary diagonal selection matrix.}\label{fig:SystemModel}
\end{figure}

Let $\mathbb{F}_2$ be the binary field. Let $\mathscr{C}_0[n,k_0]$ and $\mathscr{C}_1[n,k_1]$ be two binary linear codes of length $n$ with dimension $k_0$ and $k_1$, respectively.
We assume that $\mathscr{C}_0[n,k_0]$ has an efficient \emph{list} decoding algorithm.
Then we construct TPST code with length $2n$ and dimension $k=k_0+k_1$ as follows.
Let  $\bm{u}=(\bm{u}^{(0)},\bm{u}^{(1)})\in\mathbb{F}_2^{k}$ be a pair of information sequences to be transmitted, where $\bm{u}^{(0)}\in\mathbb{F}_2^{k_0}$ and $\bm{u}^{(1)}\in\mathbb{F}_2^{k_1}$ are associated with Layer~0 and Layer~1, respectively.
Let $\mathbf{R}$ be an $n\times n$ binary random matrix and $\mathbf{S}=\text{diag}\{s_0,\dots,s_{n-1}\}$ be a binary diagonal matrix, referred to as a \emph{selection matrix}~\cite{Wang2019spatially}.
The encoding algorithm of TPST is described in Algorithm~\ref{alg:Encode}.
Figuratively speaking~(as depicted in ~Fig.~\ref{fig:SystemModel}), the two information sequences are first encoded by the basic encoder and then superposed together in a twisted manner.
\begin{algorithm}{Twisted-pair Superposition Transmission}\label{alg:Encode}
\begin{itemize}
\item {\em Basic Encoding}: Encode $\bm{u}^{(i)}$ into $\bm{v}^{(i)} \in \mathbb{F}_2^n$ by the encoding algorithm of the basic code $\mathscr{C}_i[n,k_i]$, for $i=0,1$.
\item {\em Forward Superposition}: Compute $\bm{w}^{(0)}=\bm{v}^{(0)}\mathbf{R}$ and $\bm{c}^{(1)}=\bm{v}^{(1)}+\bm{w}^{(0)}$.
\item {\em Backward Superposition}: Compute $\bm{w}^{(1)}=\bm{c}^{(1)}\mathbf{S}$ and $\bm{c}^{(0)}=\bm{v}^{(0)}+\bm{w}^{(1)}$.
\end{itemize}
\end{algorithm}

The encoding result is a sequence $\bm{c}=(\bm{c}^{(0)},\bm{c}^{(1)}) \in \mathbb{F}_2^{2n}$.
Evidently, the TPST code is a block code constructed from the basic code, which has the coding rate $\frac{k_1+k_2}{2n}$.
Let $\mathbf{G}_i$ and $\mathbf{H}_i$ be a generator matrix and a parity-check matrix of the basic code $\mathscr{C}_i[n,k_i]$, for $i=0,1$.
The generator matrix and the parity-check matrix of a TPST code is given, respectively, by
\begin{align}
\nonumber\mathbf{G}_{\text{TPST}}
&=\left(
\begin{matrix}
\mathbf{G}_0& \\
&\mathbf{G}_1
\end{matrix}
\right)
\left(
\begin{matrix}
\mathbf{I}& \mathbf{R}\\
&\mathbf{I}
\end{matrix}
\right)
\left(
\begin{matrix}
\mathbf{I}& \\
\mathbf{S}&\mathbf{I}
\end{matrix}
\right)\\
&=\left(
\begin{matrix}
\mathbf{G}_0+\mathbf{G}_0\mathbf{RS}& \mathbf{G}_0\mathbf{R}\\
\mathbf{G}_1\mathbf{S}&\mathbf{G}_1
\end{matrix}
\right),
\end{align}
and
\begin{align}
\nonumber\mathbf{H}_{\text{TPST}}&=\left(
\begin{matrix}
\mathbf{H}_0& \\
&\mathbf{H}_1
\end{matrix}
\right)
\left(
\begin{matrix}
\mathbf{I}& \\
\mathbf{R}^T&\mathbf{I}
\end{matrix}
\right)
\left(
\begin{matrix}
\mathbf{I}&\mathbf{S}^T \\
&\mathbf{I}
\end{matrix}
\right)\\
&=\left(
\begin{matrix}
\mathbf{H}_0&\mathbf{H}_0\mathbf{S}^T\\
\mathbf{H}_1\mathbf{R}^T& \mathbf{H}_1+\mathbf{H}_1\mathbf{R}^T\mathbf{S}^T
\end{matrix}
\right).
\end{align}
We see from the construction that the generator matrix of a TPST code is decomposed into three parts: a block diagonal one corresponding to the basic codes, a block upper triangle matrix corresponding to the forward superposition, and a block lower triangle matrix corresponding to the backward superposition.
Therefore, similar to the PAC code construction, which is regarded as a form of upper-lower decomposition of the generator matrix, the TPST code construction can be regarded as a block upper-lower decomposition of the generator matrix in an alternative way.
Hence, it is expected for the TPST codes to have a good performance.

We also see that the generator matrix of a TPST code consists of three types of submatrices, the basic code generator matrices $\mathbf{G}_0$ and $\mathbf{G}_1$, the random transformation $\mathbf{R}$ and the selection matrix $\mathbf{S}$.
\begin{itemize}
\item The basic code generator matrices, $\mathbf{G}_i: \mathbb{F}_2^{k_i}\rightarrow \mathbb{F}_2^n$, transform the information sequences into basic codewords, for $i=0,1$.
The coding rate of TPST code is tunable by adjusting the information length $k_1$ and $k_2$ of the basic codes.
\item The random transformation, $\mathbf{R}: \mathbb{F}_2^n\rightarrow \mathbb{F}_2^n$, transforms the basic codeword $\bm{v}^{(0)}$ of Layer~0 into a random sequence, which is superimposed on Layer 1 at Step \emph{Forward Superposition}.
    Any error bit in Layer~0 will cause an ``avalanche effect'' on Layer~1, which is critical to distinguish the correct candidate codeword from the erroneous ones in our list decoding algorithm .
\item The selection matrix, $\mathbf{S}: \mathbb{F}_2^n\rightarrow \mathbb{F}_2^n$, transforms $\bm{c}^{(1)}$ into a masked sequence,  which is superimposed on Layer 0 at Step \emph{Backward Superposition}.
    The fraction of non-zero elements in the diagonal entries of $\mathbf{S}$, denoted by $\alpha$,  is referred to as the \emph{superposition fraction} in this paper,
    where the case $\alpha=1$  corresponds to \emph{full superposition}, and the case $\alpha<1$ corresponds to \emph{partial superposition}~\cite{Wang2019spatially}.
\end{itemize}
\subsection{Decoding}\label{sec:dec}

For simplicity, we assume that $\bm{c}$ is modulated using binary phase-shift keying~(BPSK) and 
transmitted over an additive white Gaussian noise~(AWGN) channel. 
The resulting received sequence is denoted by $\bm{y}$.
We use the notation with a hat sign $(\hat{~})$ to denote the corresponding estimated messages according to the decoder output. 

Given a received sequence $\bm{y}=(\bm{y}^{(0)},\bm{y}^{(1)})$, the optimal decoding in terms of minimizing FER is the ML decoding, which finds the TPST codeword $\bm{c}^{*}$ such that
\begin{equation}
P(\bm{y}|\bm{c}^*)=\max_{\bm{v}^{(0)}\in \mathscr{C}_0 \atop \bm{v}^{(1)}\in \mathscr{C}_1}P(\bm{y}^{(0)},\bm{y}^{(1)}|\bm{c}^{(0)},\bm{c}^{(1)}).
\end{equation}
Let $\ell_{\max}$ be a positive integer.
A sub-optimal solution is to apply the successive cancellation list decoding~(which indeed is optimal when the list size ${\ell_{\max}=2^{k_0}}$ and ML decoding is employed at Layer~1) as described below. 

1) 
From the \emph{backward superposition} as shown in Fig.~\ref{fig:SystemModel}, we have
\begin{equation}
\bm{v}^{(0)}=\bm{c}^{(0)}+\bm{c}^{(1)}\mathbf{S}.
\end{equation}
Hence, similar to polar codes~\cite{arikan2009channel}, the log-likelihood ratios~(LLRs) of $\bm{v}^{(0)}$ can be computed from $(\bm{y}^{(0)},\bm{y}^{(1)})$, the noisy version of $(\bm{c}^{(0)},\bm{c}^{(1)})$.
To be precise, by assuming that $\bm{c}^{(1)}$ is identically uniformly distributed~(i.u.d.), we compute the LLRs of $\bm{v}^{(0)}$  as

\begin{align}\label{eq:likeli_v0}
\Lambda(v_j^{(0)})=\left\{
\begin{array}{ll}
\log\frac{P(y^{(0)}_j|0)}{P(y^{(0)}_j|1)},&s_j=0\\
\log\frac{P(y^{(0)}_j|0)P(y^{(1)}_j|0)+P(y^{(0)}_j|1)P(y^{(1)}_j|1)}{P(y^{(0)}_j|1)P(y^{(1)}_j|0)+P(y^{(0)}_j|0)P(y^{(1)}_j|1)},&s_j=1
\end{array}\right.,
\end{align}
for $j=0,1,\dots n-1$, where the subscript $j$ denotes the $j$-th component of the sequence.
Taking $\Lambda(\bm{v}^{(0)})$ as input to the basic list decoder of $\mathscr{C}_0$, we obtain a list of candidate codewords $\hat{\bm{v}}_{\ell}^{(0)}, \ell=1,2,\dots,\ell_{\max}$.

2) 
From the encoding process of TPST, we have
\begin{equation}\label{eq:v1}
(\bm{v}^{(1)}\mathbf{S},\bm{v}^{(1)})=(\bm{c}^{(0)},\bm{c}^{(1)})+\bm{v}^{(0)}(\mathbf{I}+\mathbf{RS},\mathbf{R}),
\end{equation}
indicating that $\bm{v}^{(1)}$ is transmitted twice~(with partial masking if partial superposition is employed).
Hence, if $\bm{v}^{(0)}$ is obtained by the decoder, the LLRs of $\bm{v}^{(1)}$ can be computed as,
\begin{align}\label{eq:likeli_v1}
\nonumber\Lambda(v^{(1)}_j)=&
\log\frac{P(y^{(1)}_j|w_{j}^{(0)})}{P(y^{(1)}_j|w_{j}^{(0)}+1)}\\
&+s_j \log\frac{P(y^{(0)}_j|w_{j}^{(0)}+v_{j}^{(0)})}{P(y^{(0)}_j|w_{j}^{(0)}+v_{j}^{(0)}+1)},
\end{align}
for $j=0,1,\dots n-1$.
However, since $\bm{v}^{(0)}$ is unknown at the receiver, we calculate $\Lambda_{\ell}(\bm{v}^{(1)})$ instead by treating each candidate codeword $\hat{\bm{v}}_{\ell}^{(0)}$ as correct.
Then we can estimate $\hat{\bm{v}}_{\ell}^{(1)}$ by taking $\Lambda_{\ell}(\bm{v}^{(1)})$ as input to the basic decoder of $\mathscr{C}_1$.

We see that the LLRs $\Lambda_{\ell}(\bm{v}^{(1)})$ calculated by~(\ref{eq:likeli_v1}) match with the channel if and only if the candidate $\hat{\bm{v}}_{\ell}^{(0)}$ is correct.
In the case when $\hat{\bm{v}}_{\ell}^{(0)}$ is erroneous, the mismatch can be roughly measured by the weight of binary interference $\Delta\bm{v}(\mathbf{I}+\mathbf{RS},\mathbf{R})$, where $\Delta\bm{v}=\hat{\bm{v}}_{\ell}^{(0)}+\bm{v}^{(0)}\neq\bm{0}$.
Since $\mathbf{R}$ is randomly generated, a nonzero $\Delta\bm{v}$ can cause a significant change on the joint typicality between $(\bm{c}_{\ell}^{(0)},\bm{c}_{\ell}^{(1)})$ and $(\bm{y}^{(0)},\bm{y}^{(1)})$.
This can be quantified by EDF in Section~\ref{sec:t} as defined in~\cite{ma2019statistical}.

For each candidate pair $\hat{\bm{v}}_{\ell}=(\hat{\bm{v}}_{\ell}^{(0)},\hat{\bm{v}}_{\ell}^{(1)})$, the decoder computes the likelihood of the corresponding TPST codeword $P(\bm{y}|\hat{\bm{c}}_{\ell})$ and selects the most likely one as the decoding output.

The decoding procedure is summarized in Algorithm~\ref{alg:Decode}.
\begin{algorithm}{Successive Cancellation List Decoding of TPST}\label{alg:Decode}
\begin{itemize}
\item {\em Initialization}: Take the received sequence $\bm{y}$ as input and compute the likelihood of $\bm{v}^{(0)}$ given by~(\ref{eq:likeli_v0}).
\item {\em List Decoding}: For $1 \leqslant \ell \leqslant \ell_{\max}$,
\begin{enumerate}
\item[1)] Employ the list decoding algorithm of the basic code, resulting in the $\ell$-th list decoding output $\hat{\bm{v}}_{\ell}^{(0)}$.
\item[2)] Compute the LLRs of $\bm{v}^{(1)}$ given by~(\ref{eq:likeli_v1}).
\item[3)] Employ the decoding algorithm of the basic code, resulting in the decoding output $\hat{\bm{v}}_{\ell}^{(1)}$.
\item[4)] Compute the likelihood of the TPST codeword $P(\bm{y}|\hat{\bm{c}}_{\ell})$.
\end{enumerate}
\item {\em Output}:  Output the TPST codeword $\hat{\bm{c}}$ such that
$$P(\bm{y}|\hat{\bm{c}})=\max_{1 \leqslant \ell \leqslant \ell_{\max}}P(\bm{y}|\hat{\bm{c}}_{\ell}).$$
\end{itemize}
\end{algorithm}

\section{Performance Analysis}\label{sec:performance}
\subsection{Performance Bounds}
To analyze the performance of the proposed SCL decoding algorithm for TPST codes, we consider the following events.
\begin{itemize}
\item The event $E_0$ that the transmitted basic codeword $\bm{v}^{(0)}$ of Layer~0 is not in the list, i.e., $\bm{v}_{\ell}^{(0)}\neq \bm{v}^{(0)}$ for $\ell =1,2,\dots,\ell_{\max}$;
\item The event $E_1$  that given the transmitted basic codeword $\bm{v}^{(0)}$ of Layer~0, the basic decoder does not output the transmitted basic codeword $\bm{v}^{(1)}$ of Layer~1, i.e., $\bm{v}_{\ell}^{(1)}\neq \bm{v}^{(1)}$ if $\bm{v}_{\ell}^{(0)} = \bm{v}^{(0)}$;
\item The event $E_2$ that there exists a valid codeword in the decoding list which is more likely than the transmitted codeword, i.e., $P(\bm{y}|\hat{\bm{c}}_{\ell})>  P(\bm{y}|\bm{c})$ for some $\ell \in\{1,2,\dots,\ell_{\max}\}$.
\end{itemize}

Hence, the FER performance of the TPST code with the proposed SCL decoding can be expressed as
\begin{align}\label{eq:fer}
\text{FER}_{\text{SCL}}=P(E_0\cup E_1\cup E_2).
\end{align}
Then we have the following lower bounds.

1) Genie-aided bound $P(E_0)$ for Layer~0

The received vector $\bm{y}^{(0)}$ can be viewed as a noisy version of $\bm{v}^{(0)}$, where $\bm{c}^{(1)}$ is considered as binary interference with side information $\bm{y}^{(1)}$.
The link $\bm{v}^{(0)}\rightarrow\bm{y}^{(0)}$ is referred to as the \emph{binary interference AWGN channel}.
From~(\ref{eq:fer}),  we see that $P(E_0)$ is a lower bound of $\text{FER}_{\text{SCL}}$.
To estimate $P(E_0)$, we consider a genie-aided list decoder in which the decoder outputs the transmitted codeword if it is in the list~(told by a genie).
Hence, the genie-aided bound $P(E_0)$  can be obtained by simulations on the list decoding performance of the basic code for Layer~0 over the binary interference AWGN channel.

2) Genie-aided bound $P(E_1)$ for Layer~1

By removing the effect of $\bm{v}^{(0)}$, both the received vector $\bm{y}^{(0)}$~(or only parts of $\bm{y}^{(0)}$ when partial superposition is employed) and $\bm{y}^{(1)}$ can be equivalently viewed as noisy versions of $\bm{v}^{(1)}$ transmitting over the AWGN channel.
The link $\bm{v}^{(1)}\rightarrow\{(\bm{y}^{(0)},\bm{y}^{(1)})|\bm{v}^{(0)}\}$ is referred to as the \emph{repetition AWGN channel}.
Obviously, $P(E_1)$ is also a lower bound of $\text{FER}_{\text{SCL}}$.
To estimate $P(E_1)$, we consider a genie-aided decoder for decoding $\bm{v}_{\ell}^{(1)}$ in which the decoder is told by a genie that the correct information bits $\bm{v}_{\ell}^{(0)}$.
Hence, $P(E_1)$  can be obtained by by simulations on the performance of the basic code for Layer~1 over the repetition AWGN channel.

\begin{figure}[t]
  \centering
  \includegraphics[width=\figwidth]{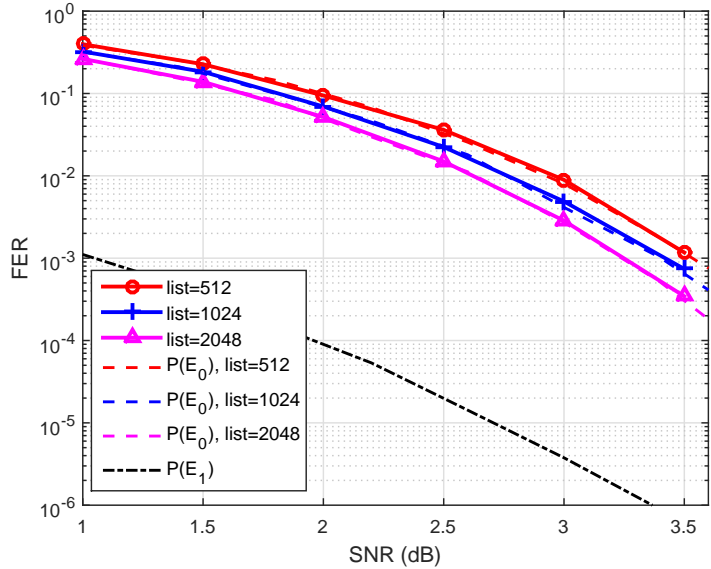}\\
  \caption{Genie-aided bounds of TPST-TBCCs with different list sizes. The dashed curves are genie-aided bounds for Layer~0, the dash-dotted curve is the genie-aided bound for Layer~1, and the solid curves are the simulation performance.}\label{FIG_LIST}
\end{figure}

{\bf Remark.} Notice that both $P(E_0)$ and $P(E_1)$ are irrelevant to the random transformation $\mathbf{R}$. This makes it more convenient to analyze the performance by the lower bound given by
\begin{equation}\label{eq:genie}
\text{FER}_{\text{SCL}}\geqslant \max\{P(E_0),P(E_1)\}.
\end{equation}
\begin{example}\label{exp:list}
We take the TBCCs as basic codes, which can be efficiently list decoded by adopting the list Viterbi algorithms, resulting in the TPST-TBCCs.
First we take the $(2,1,4)$ TBCC defined by the polynomial generator matrix $G(D) = (1+D^2+D^3+D^4, 1+D+D^4)$~(also denoted in octal notation as $(56,62)_8$) with information length $32$ as the basic codes for both the two layers.
We sample the random matrix $\mathbf{R}$ from  the permutation matrices and let $\mathbf{S}$ be the identity matrix for simplicity.
As a result, the constructed TPST-TBCC is a rate-$1/2$ block code with length $128$. The genie-aided bounds are shown in Fig.~\ref{FIG_LIST}, where the simulated error rates of Layer~0 are also plotted. We can observe that
\begin{itemize}
  \item The genie-aided performance of $\bm{v}^{(0)}$ can be improved by increasing the list size $\ell_{\max}$, as expected.
  \item The curves of simulated error rate of $\bm{v}^{(0)}$ match very well with the genie-aided bounds, which indicates that the decoder can well distinguish the correct codeword~(once it is in the decoding list) from erroneous ones.
      Hence we can use~(\ref{eq:genie}) to estimate the decoding performance of the TPST-TBCC.
      The matched performance between the simulation and the genie-aided bound $P(E_0)$ also indicates that the use of random interleaver instead of random transformation only causes a negligible performance loss. Evidently, this reduces much the encoding/decoding complexity for practical implementations.
  \item When taking the same basic code for the two layers and employing full superposition, there exists a large gap between the genie-aided performance of $\bm{v}^{(0)}$ and $\bm{v}^{(1)}$. As a result, the decoding of $\bm{v}_{\ell}^{(0)}$ dominates FER performance of the TPST code. 
\end{itemize}
\end{example}

3) Lower bound $P(E_2)$ for ML decoding

Let $\text{FER}_{\text{ML}}$ be the FER performance of the TPST code with ML decoding.
Since the ML decoding outputs the most likely codeword,  an error occurs if and only if the transmitted codeword $\bm{c}$ is not with the highest likelihood among the codebook $\mathscr{C}$.
Therefore, once the given decoder outputs a valid codeword that is more likely than the transmitted one,  the ML decoding will surely make an error in this instance as well.
Hence, $P(E_2)$ is not only a lower bound of $\text{FER}_{\text{SCL}}$, but also a lower bound of $\text{FER}_{\text{ML}}$.
We can estimate $P(E_2)$ by simulating the frequency of the event that the decoding output has  a higher likelihood than the transmitted one.
Then the performance of ML decoding is bounded by
\begin{equation}
P(E_2)\leqslant \text{FER}_{\text{ML}}\leqslant \text{FER}_{\text{SCL}}.
\end{equation}

Now, we show that the proposed SCL decoding attains a near-ML performance as follows.
By applying the union bound technique, we obtain an upper bound of $\text{FER}_{\text{SCL}}$ from~(\ref{eq:fer}) as
\begin{equation}
\text{FER}_{\text{SCL}} \leqslant P(E_0)+P(E_1)+P(E_2).
\end{equation}
Therefore, the gap between the SCL performance and the ML performance can be upper bounded by
\begin{equation}\label{eq:mlgap}
\text{FER}_{\text{SCL}} - \text{FER}_{\text{ML}} \leqslant \text{FER}_{\text{SCL}}-P(E_2)  \leqslant  P(E_0)+P(E_1).
\end{equation}
Hence, by lowering down the genie-aided bounds for the two layers, we can construct a TPST code with an SCL decoding algorithm attaining a near-ML performance.

More specifically, if the basic decoder for Layer 2 is ML decoding~(e.g., taking TBCC with Viterbi decoding algorithm as basic code), the upper bound of the gap can be further tightened.
In this case, the instance that $\bm{v}^{(0)}$ of Layer~0 is in the list and the basic decoder of Layer~1 outputs a basic codeword  $\hat{\bm{v}}^{(1)}$ of Layer~1 other than $\bm{v}^{(1)}$ of Layer~1 implies that the SCL decoder finds a valid TPST codeword more likely than the transmitted one.
Hence, we have $E_1 \subseteq E_2$ and
\begin{equation}
\text{FER}_{\text{SCL}} =P(E_0 \cup E_2)  \leqslant  P(E_0)+P(E_2).
\end{equation}
Then the gap between the SCL performance and the ML performance is upper bounded by
\begin{equation}\label{eq:mlgap2}
\text{FER}_{\text{SCL}}- \text{FER}_{\text{ML}}   \leqslant \text{FER}_{\text{SCL}}-P(E_2)  \leqslant  P(E_0),
\end{equation}
the genie-aided bound for Layer~0.
\subsection{Complexity Analysis and Early Termination}\label{sec:t}

We see from Algorithm~\ref{alg:Decode} that, when taking convolutional codes as basic codes, the decoding complexity is roughly  $\ell_{\max}$ times as much as that of the Viterbi algorithm for both the basic codes of the two layers.
To attain a near-ML performance, the decoding requires a large list size $\ell_{\max}$, which incurs a high decoding complexity.
However, the average list size for $\bm{v}^{(0)}$ to be included in the decoding list is typically small, especially in the high SNR region.
Hence, to reduce the complexity, we design a criterion for identifying the correct codeword, so that the list decoding can be early terminated.

We use the metric of \emph{empirical divergence function}~(EDF) defined in~\cite{ma2019statistical} for the early termination, which is given as
\begin{equation}
D(\bm{y},\bm{v})=\frac{1}{2n}\log_2\frac{P(\bm{y}|\bm{c})}{P(\bm{y})},
\end{equation}
where $P(\bm{y})$ is obtained by assuming a uniformly distributed input.
As illustrated in Section~\ref{sec:dec}, an erroneous candidate can cause a significant change on the joint typicality between $(\hat{\bm{c}}_{\ell}^{(0)},\hat{\bm{c}}_{\ell}^{(1)})$ and $(\bm{y}^{(0)},\bm{y}^{(1)})$.
Hence, the EDF associated with an erroneous candidate  is typically less than that of the correct one.
Similar to~\cite{ma2019statistical}, we set an off-line learned threshold $T$ for early termination, where the decoding candidate $\hat{\bm{v}}_{\ell}=(\hat{\bm{v}}_{\ell}^{(0)},\hat{\bm{v}}_{\ell}^{(1)})$ is treated as correct if $D(\bm{y},\hat{\bm{v}}_{\ell}) > T$.

\section{Design for Finite-length Regime}\label{sec:design}
As discussed above, we see that the performance of a TPST code is lower bounded by $\max\{P(E_0),P(E_1)\}$, and the gap to the ML decoding is upper bounded by $P(E_0)+P(E_1)$.
To jointly optimize these two bounds, we expect that both the genie-aided bounds, $P(E_0)$ and $P(E_1)$, simultaneously attain the target error rate.
However, we see from Example~\ref{exp:list} that, there exists a gap between the genie-aided bounds of the two layers, $P(E_0)\gg P(E_1)$, resulting in poor performance.
To improve the performance, we consider two approaches to narrowing the gap. One is to reduce the coding rate of Layer~0 by changing the basic codes with different rates, and the other is to improve the channel for Layer~0 by introducing partial superposition at Step \emph{Backward Superposition} to reduce the binary interference from Layer~1.
\subsection{Rate Allocation}\label{rate_allo}
From the derivation of the genie-aided lower bounds, we see that when full superposition is employed, the genie-aided bounds can be obtained by simulations on the basic codes transmitted over binary interference AWGN channels and the repetition AWGN channels.
Based on the genie-aided bounds, we can choose the basic codes with different rates to construct a good TPST code with a target error rate.

Suppose that we have a family of basic codes of length $n$ with different rates with their genie-aided bounds available.
For example, the basic codes with different rates can be constructed by puncturing from a mother code and the genie-aided bounds can be obtained by off-line simulations over the binary interference AWGN channel and the repetition AWGN channel.
Then the procedure of designing TPST code with length $2n$ and dimension $k$  by the rate allocation approach is shown as follows.
\begin{enumerate}
\item According to the error performance over the repetition AWGN channel, determine $k_1$ such that a  basic code with dimension $k_1$ has an error performance near the target error rate, which is taken as $\mathscr{C}_1[n,k_1]$, the basic code for Layer~1.
\item Let $k_0=k-k_1$. Take the basic code with dimension $k_0$ as $\mathscr{C}_0[n,k_0]$, the basic code of Layer~0.
\item Increase the decoding list size $\ell_{\max}$ of $\mathscr{C}_0[n,k_0]$ such that the list decoding performance of $\mathscr{C}_0[n,k_0]$  over the binary interference AWGN channel achieves an error rate near the target one.
\end{enumerate}

\begin{figure}[t]
  \centering
  \includegraphics[width=\figwidth]{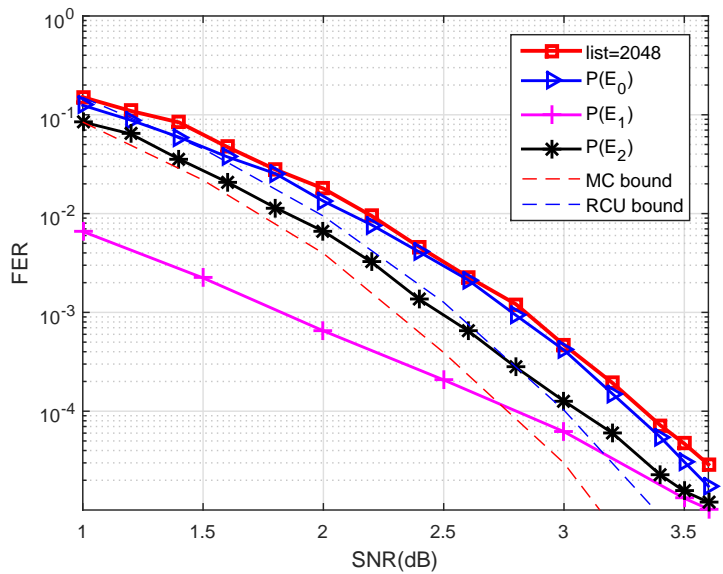}\\
  \caption{Decoding performance of the TPST-TBCC designed by the rate allocation approach.}\label{FIG_RATBCC}
\end{figure}

\begin{example}
We consider random codes with ordered-statistic list decoding as basic codes for Layer~0 and (punctured)~TBCCs with Viterbi decoding as basic codes for Layer~1.
To construct a TPST code with length $128$ and dimension $64$ targeting on the FER at the level of $10^{-5}$, we use the (punctured) $(2,1,4)$ TBCC defined by the polynomial generator matrix $G(D) = (56,62)_8$ to construct the basic codes with different rates.
The genie-aided bounds of the basic codes are shown in Fig.~\ref{FIG_RAOSD}.
We can observe that
\begin{itemize}
\item As the increase of the dimension $k_1$ of the TBCC codes the performance of genie-aided bounds for Layer~1 decreases. Targeting on the FER at the level of $10^{-5}$ and comparing with the RCU bound, we choose the code with dimension $k_1=35$ as the basic code for Layer~1, which achieves the target FER performance at the SNR about 3.6~dB.
\item Then we take random code with length $64$ and dimension $k_0=29$ with OSD as the basic code for Layer~0. To attain the targeting FER performance, we consider the OSD algorithm with order $5$ for the decoding of  Layer~0.
    However, it suffers from an extremely high decoding complexity~(where the list size is $146596$ for OSD with order $5$) and hence is impractical for implementation.
\end{itemize}
\begin{figure}[t]
  \centering
  \includegraphics[width=\figwidth]{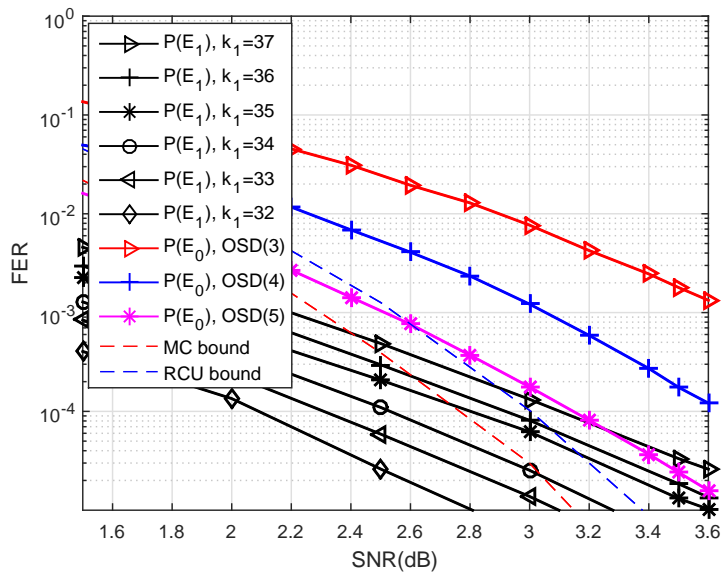}\\
  \caption{Genie-aided bounds of the basic codes. For Layer~1, the basic codes are (punctured) $(2,1,4)$ TBCCs defined by the polynomial generator matrix $G(D) = (56,62)_8$ with length $64$ and dimension from $32$ to $37$. For Layer~0, the basic code is the random code with length $64$ and dimension $29$.}\label{FIG_RAOSD}
\end{figure}

Alternatively, we consider TBCC with list Viterbi algorithm as the basic code for Layer~0.
We puncture the $(3,1,4)$ TBCC defined by the polynomial generator matrix $G(D) = (52, 66, 76)_8$ to fit the length $64$ and dimension $29$, where the performance is shown in Fig.~\ref{FIG_RATBCC}.
We can observe that
\begin{itemize}
\item  By setting a maximum list size $\ell_{\max}=2048$, the genie-aided bound for Layer~0 has a similar performance to that of Layer~1 at the SNR about $3.5$~dB.
\item  The constructed TPST code has an ML performance near the RCU bound for finite length codes. Also, the SCL decoding of the constructed TPST code has a near-ML performance, which is within $0.4$~dB away from the RCU bound at the FER of $10^{-5}$.
\end{itemize}
\end{example}

We see from the above examples that the list Viterbi decoding of TBCC is more efficient than the OSD decoding in the sense of list size. Hence we consider TBCC as basic codes for the TPST code construction in the sequel.
\subsection{Partial Superposition}\label{sec_part_sup}
Different from the rate allocation approach that adjusts the coding rate of the two layers of TPST codes, the following partial superposition approach adjusts the  channels for the two layers.
We introduce the so-called superposition fraction $\alpha$ to characterize the partial superposition.
Given $\alpha$, we construct the binary diagonal matrix $\mathbf{S}$ by assigning $\lfloor n\alpha \rfloor$ ones and ${n-\lfloor n\alpha \rfloor}$ zeros homogeneously~(as uniformly as possible) to the $n$ positions.
For simplicity of designing, we employ the same basic code on both layers.

1) Recall that the link $\bm{v}^{(0)}\rightarrow\bm{y}^{(0)}$ can be viewed as the binary interference AWGN channel, where $\bm{c}^{(1)}$ is considered as binary interference with side information $\bm{y}^{(1)}$.
Therefore, $P(E_0)$ can be lowered down by reducing the binary interference, which can be achieved by decreasing the superposition fraction $\alpha$.

2) Recall that with $\bm{v}^{(0)}$ available, the link $\bm{v}^{(1)}\rightarrow\{(\bm{y}^{(0)},\bm{y}^{(1)})|\bm{v}^{(0)}\}$ can be viewed as the repetition AWGN channel, where $\bm{v}^{(1)}$ is transmitted twice over AWGN channels~(one is with partial masking).
Therefore, decreasing the superposition fraction $\alpha$ leads to performance degradation of the genie-aided decoder for $\bm{v}^{(1)}$ and hence increases $P(E_1)$.

In summary, compared with full superposition, decreasing the superposition fraction $\alpha$ can narrow the gap between the two genie-aided bounds, as illustrated in Example~\ref{exp:frac} below.

\begin{figure}[t]
  \centering
  \includegraphics[width=\figwidth]{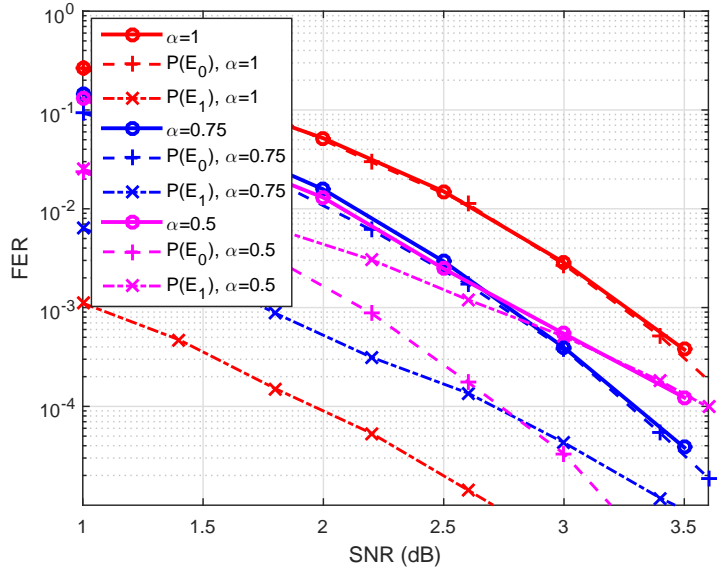}\\
  \caption{Decoding performance of the TPST-TBCCs with different superposition fractions. The dashed curves are genie-aided bounds for Layer~0, the dash-dotted curves are genie-aided bounds for Layer~1, and the solid curves are the simulation performance.}\label{FIG_FRAC}
\end{figure}

\begin{example}\label{exp:frac}
We take the same basic code as in Example~\ref{exp:list} and construct the TPST code with different superposition fractions $\alpha =0.5, 0.75~\text{and}~1.0$. The genie-aided bounds for $\ell_{\max}=2048$ are shown in Fig.~\ref{FIG_FRAC}.
We can observe that a smaller superposition fractions $\alpha$ results in a lower genie-aided bound for decoding Layer~0 and a higher genie-aided bound for decoding Layer~1, which verifies our analysis.
\end{example}

Hence, we construct TPST codes by optimizing the superposition fraction $\alpha$ such that the genie-aided bounds of the two layers have a similar performance in the SNR region of interest.
\section{Numerical Simulation}\label{sec:simulation}
The flexibility of the construction lies in the following facts.
1) Any block codes with a fast encoding and efficient list decoding can be adopted as basic codes.
2) Given a family of basic codes with different rates, it is easy to construct good TPST codes by the rate allocation approach for a target FER.
3) For the basic codes with a given coding rate, the construction can be easily improved by optimizing the superposition fraction.
In this subsection, we take TPST-TBCCs as examples and show by numerical simulations that the constructed TPST-TBCCs, with a wide range of coding rates, have near-capacity performance in the finite length regime.
\subsection{Construction with Different Rates}

\begin{figure}[t]
  \centering
  \includegraphics[width=\figwidth]{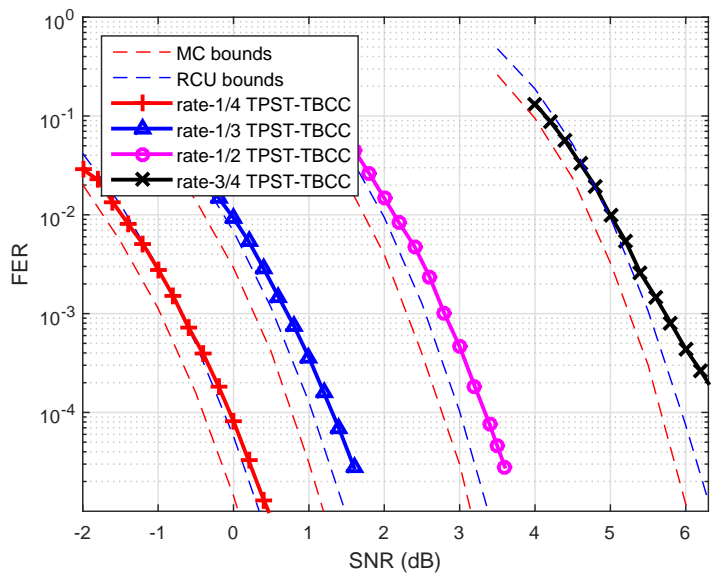}\\
  \caption{Decoding performance of the TPST-TBCCs with different coding rates.}\label{FIG_RATE}
\end{figure}
\begin{figure}[t]
  \centering
  \includegraphics[width=\figwidth]{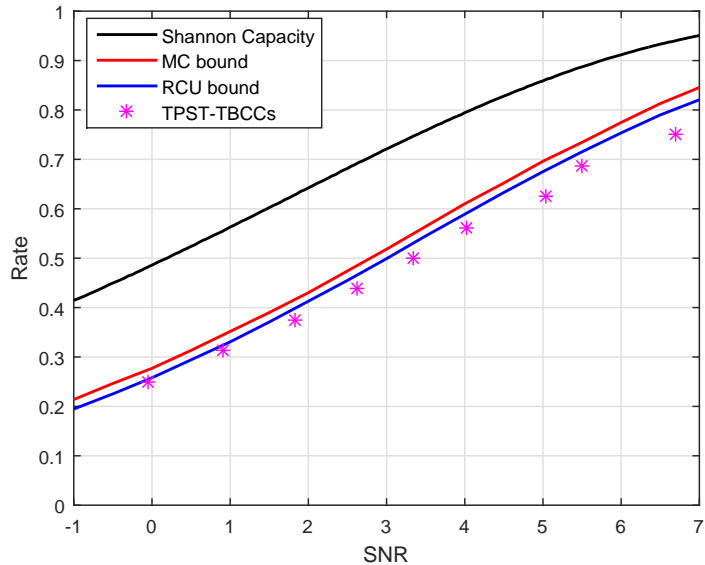}\\
  \caption{Performance of TPST-TBCCs with length $128$ at the FER of $10^{-4}$.}\label{FIG_CA}
\end{figure}

\begin{example}\label{exp:rate}
We use TBCCs with different coding rates presented in~\cite{stahl1999optimal} as the mother codes for the basic codes, where their generators~(in octal notation) are listed in Table~\ref{TAB_RATE}.
For example, to construct a basic codeword with coding rate $3/4$, we puncture every third bit of the TBCC $\mathscr{C}[96,48]$ encoder output.
The TPST codes are encoded with superposition fraction $\alpha =0.75$ and decoded with list size $\ell_{\max}=2048$.  The simulation results are shown in Fig.~\ref{FIG_RATE}.
For comparison, we have also plotted the coding bounds on the minimum FER in the finite length regime. We can observe that the proposed TSPT-TBCCs can well approach the finite length capacity with different rates.
Focusing on the performance at the FER of $10^{-4}$, the comparison between the performance of TPST-TBCCs and channel capacity is shown in Fig.~\ref{FIG_CA}.
We see that the TPST-TBCCs with a lower coding rate can better approach to the RCU bound. This is probably because the basic code with a lower coding rate has a smaller codebook size, resulting in a relatively better list decoding performance when the list size is fixed.
\end{example}

\begin{table}\caption{Basic TBCCs Used in Example~\ref{exp:rate}}
\centering
  \begin{tabular}{|c|c|c|}\hline
  Rate  & Memory & Generators\\\hline
   1/4  &  4  & $(52, 56, 66, 76)_8$  \\\hline
   1/3  &  4  & $(52, 66, 76)_8$ \\\hline
   1/2  &  4  & $(56, 62)_8$  \\\hline
  \end{tabular}\label{TAB_RATE}
\end{table}

\subsection{Tradeoff between Complexity and Performance}
\begin{figure}[t]
  \centering
  \includegraphics[width=\figwidth]{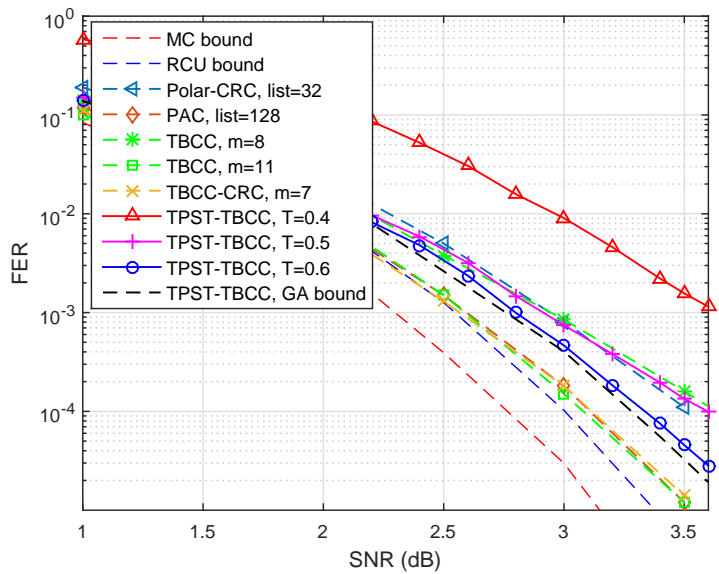}\\
  \caption{Decoding performance of the TPST-TBCCs with different thresholds.}\label{FIG_T}
\end{figure}

\begin{example}\label{exp:t}
We take the same basic code as in Example~\ref{exp:list}. The TPST code is encoded with superposition fraction $\alpha =0.75$ and decoded with list size $\ell_{\max}=2048$, where different thresholds are used for early termination.  The simulation results are shown in Fig.~\ref{FIG_T}, and their corresponding decoding complexity in terms of average list size are given in Table~\ref{TAB_T}. For comparison, we have also redrawn the performance curves of polar-CRC code and PAC code summarized in~\cite{Yao2020list}, TBCCs summarized in~\cite{coskun2019efficient} and TBCC-CRC presented in~\cite{Liang2019list} with length $n=128$ and dimension $k=64$. We can observe that
\begin{itemize}
  \item The complexity~(average list size), at the cost of performance loss, can be reduced by tuning down the threshold $T$.
  \item The proposed TPST-TBCC with list decoding has a competitive performance with the TBCCs with large memory. However, in our TPST construction, the basic TBCCs are with memory $m=4$ which have a much smaller number of states in their trellises and hence have lower decoding complexity than that of the TBCCs with large memory.
  \item By selecting properly a threshold, the decoding of TPST code has a near-capacity performance with a relatively low decoding complexity. For example, at the SNR of $3.6$~dB, the FER performance of the presented TPST-TBCC with threshold $T=0.5$ is about $0.6$~dB away from the RCU bound. Whereas, the average list size for the SCL decoding is only around two, which is much less than the maximum list size 2048.
\end{itemize}
\end{example}
\begin{table}\caption{Average List Size Needed for Different Threshold $T$}
\centering
  \begin{tabular}{l|l|l|l|l|l|l|l|l}\hline
  $\text{SNR}$           & 1.0 & 1.4 & 1.8 & 2.2 & 2.6 & 3.0& 3.4&3.6\\\hline
   $T=0.4$     & 12.1  & 18.1  & 18.8  & 13.1  & 9.1  & 3.7& 1.8&1.4\\\hline
   $T=0.5$     & 459  & 275  & 132 & 55.3 & 22.9 & 7.5& 2.7&1.9\\\hline
   $T=0.6$     & 1412  & 1042  & 685 & 396 & 199 & 86.6& 33.4&20.1\\\hline
  \end{tabular}\label{TAB_T}
\end{table}

\section{Conclusion}\label{sec:conclusion}
In this paper, we propose a novel coding scheme referred to as twisted-pair superposition transmission, which is constructed by mixing up the basic codes by superposition.
Based on the list decoding of the basic codes, we present an SCL decoding algorithm for the TPST codes.
To reduce the complexity, we present thresholds for early termination of the decoding, which can significantly reduce the decoding complexity at the expense of a slight decoding performance degradation.
We derive bounds for the FER performance, which shows that the performance of the proposed SCL decoding is near-ML and can be easily predicted by the genie-aided bounds obtained from the basic codes.
Based on the genie-aided bounds, we have present two design approaches for the construction of TPST codes.
Taking TBCCs as basic codes, we show by numerical simulations that the constructed TPST-TBCCs have near-capacity performance in a wide range of coding rates in the short length regime.
Hence, we expect TPST codes to be potentially applicable to future low latency communications.
\bibliographystyle{IEEEtran}
\bibliography{IEEEabrv,TPST-TCOM}


%
%
%
%
%
%
%

\end{document}